# Agent Operating Systems (AOS): Integrating Agentic Control Planes into, and Beyond, Traditional Operating Systems

Ankur Sharma
Independent Researcher
San Francisco, USA
ankur.sharma@ocproject.net

Deep Shah
Independent Researcher
deepshah146@gmail.com

## Abstract

Traditional operating systems were designed around deterministic programs, explicit control flow, and human initiated workflows. Their core abstractions processes, threads, system calls, files, and permissions assume bounded behavior and predictable interaction patterns. Agentic AI systems introduce a different execution model: long-lived, goal-directed entities that reason probabilistically, invoke tools dynamically, and adapt behavior based on feedback. While agents can be implemented as user-space applications today, their execution characteristics stress OS boundaries in scheduling, memory and state management, security, observability, and governance.

This paper introduces the concept of an Agent Operating System (AOS), a systems architecture that integrates an agentic control plane into existing operating systems or, in some models, subsumes selected OS responsibilities over time. We provide a precise definition of an AOS, explicit assumptions and non-goals, and a structured decomposition of AOS responsibilities into schedulers, context and memory management, tool and capability registries, policy and trust enforcement, and observability and audit. We analyze limitations of classical OS abstractions for agent workloads, propose integration models from user-space runtimes to distributed control planes, and map AOS concepts onto Linux and Windows primitives. We present security and safety implications, including agent specific threat models, and define evaluation criteria that emphasize deterministic enforcement, auditability, and operator comprehensibility. The objective is not to replace operating systems wholesale, but to establish a rigorous systems foundation for agentic computation that remains controllable, accountable, and secure at scale.

## 1. Introduction and Motivation

Operating systems represent one of the most successful abstractions in computing because they provide stable contracts between hardware, applications, and operators. These contracts assume a specific execution model: deterministic programs invoked by users or services, with explicit control flow and explicit control of I/O. The OS provides concurrency primitives, memory and isolation, file and device abstractions, and security boundaries. While modern OSes support containers and virtual machines, their foundational units remain processes, threads, and system calls.

Agentic systems introduce a workload class that does not fit cleanly into this model. An agent is not merely an interactive program or a script. It is a long-lived control loop that maintains internal state, performs probabilistic reasoning, selects actions opportunistically, and adapts based on feedback signals. Its action selection is not a direct function of a program counter and deterministic branching alone. Instead, it may involve inference over context, retrieved memory, and uncertain environment state. Agents may dynamically choose tools, sequence tool calls, revise plans, or suspend and resume with new constraints.

Current practice often treats agents as applications that happen to call APIs. This is a reasonable early engineering approach, but it creates recurring gaps in production systems. Resource scheduling becomes misaligned because agent progress depends on external events, tool latencies, and budget constraints. Memory is mischaracterized because agent state is not only heap memory; it includes contextual memory, retrieved knowledge, and provenance metadata with distinct lifetimes and trust requirements. Security is incomplete because agents frequently require scoped permissions to call tools that cause side effects. Observability is insufficient because operators need decision lineage, tool invocation traces, and policy compliance evidence, not only CPU and memory metrics. Governance is missing because safety and policy enforcement must be deterministic even if reasoning is nondeterministic.

This paper frames a systems question: if agents become first-class computational entities, what should the operating system boundary look like? Should the OS remain unchanged, and agents live entirely in user-space? Should the OS expose new primitives for agent scheduling, memory, and governance? Or should an agentic control plane evolve to subsume parts of OS responsibility similarly to how cluster schedulers and orchestration systems subsumed higher-layer machine management responsibilities?

We propose an Agent Operating System (AOS) as a structured answer. An AOS is not a claim that existing OSes are obsolete. It is a proposal to formalize the components and boundaries needed to run agents reliably and safely on real systems while preserving determinism at the enforcement boundary.

AOS does not replace kernels; it introduces intent and governance as first-class scheduling and enforcement concepts above them. In this sense, AOS represents a semantic elevation of control rather than a reimplementation of operating systems. The research challenge is not in making agents autonomous, but in making autonomy deterministic at the boundary.

### 1.1 Assumptions:

1. LLMs and learning components are probabilistic, may be incorrect, and may be adversarially influenced.
2. The system boundary must be deterministic in enforcement, audit, and safety decisions.
3. Agent reasoning must be separable from execution and side effects.
4. Real systems constraints apply latency, throughput, cost, operator simplicity, and failure modes.

### 1.2 Non-goals:

1. Artificial general intelligence or self-improving autonomous kernels.
2. A requirement that the kernel runs model inference.
3. A claim that classical OS abstractions should be discarded wholesale.
4. A universal agent framework or application layer.

### 1.3 Contributions:

1. A precise definition of an Agent Operating System with explicit boundaries.
2. A decomposition of AOS responsibilities into lifecycle, scheduling, memory, tool mediation, policy, and observability.
3. Integration models ranging from user-space runtime to distributed control plane to partial subsumption of OS functions.

4. A mapping of AOS concepts onto Linux and Windows primitives and pragmatic migration paths.
5. Security and governance analysis with agent specific threat models and deterministic enforcement requirements.
6. Evaluation criteria and open research problems for building credible AOS implementations.

This work was conducted as part of the Unified Intelligent Infrastructure workstream at the Open Compute Project (OCP).[1]

## 2. Limitations of Classical OS Abstractions for Agents

This section identifies concrete mismatches between classical OS design assumptions and agent workload characteristics. The objective is not to argue that existing OS abstractions are incorrect. The objective is to show where they are incomplete for goal-directed, tool using entities and where ad hoc user-space implementations tend to fragment enforcement and observability.

### 2.1 Process and Thread Abstractions Assume Bounded Execution

A process typically represents a program instance with a clear start, a clear termination, and a bounded working set. Even long-lived daemons and services usually have stable responsibilities, stable APIs, and well understood call patterns. Agents frequently violate these assumptions.

First, agents are long-lived, stateful loops. They may remain resident for days or weeks, evolving state in ways not captured by process memory alone. Second, agents activate opportunistically. They may remain idle awaiting external triggers, then burst into activity to plan, call tools, and coordinate. Third, agent tasks span multiple tool invocations and external systems, where the process is only one component of the overall lifecycle and may be restarted or migrated while the agent identity persists.

Treating an agent as a process is workable but incomplete. The process abstraction cannot represent intent, goal progress, delegation context, or policy compliance. It primarily represents execution context and resource usage.

### 2.2 Scheduling Assumes Instruction Progress, Not Goal Progress

OS schedulers allocate CPU time slices and manage blocking on I/O. Agent progress depends on a broader set of resources and constraints.

Agents may spend most of their time waiting for tools, remote services, or human input. Their actions may have cost budgets, risk budgets, or rate limits, requiring scheduling decisions based on budget utilization rather than CPU time alone. Agents also require dependency scheduling: they might need to coordinate sub tasks, wait for evidence, or reconcile conflicting signals before proceeding.

In classical scheduling, a blocked thread is often treated as idle from a CPU viewpoint. In agent systems, blocked time may represent pending tool calls, pending approvals, pending policy checks, or external event dependencies. The scheduler must interpret these states as part of goal progress, not as mere inactivity.

### 2.3 Memory Is Not Merely Addressable Storage

OS memory management allocates pages; controls address spaces and ensures isolation. Agents require additional memory semantics.

[1] https://www.opencompute.org/w/index.php?title=Unified_Intelligent_Infrastructure

Agent state includes short term context windows, long term episodic state, retrieved knowledge, and provenance metadata with different lifetimes and trust levels. Operational requirements frequently depend on replay semantics: the ability to reconstruct why a decision was taken using the same memory view, even if the model is probabilistic.

Classical memory abstractions do not provide structured lifetimes, trust labels, or provenance at the level required for audits. As a result, agent memory is implemented ad hoc in application layers, fragmenting security and observability and producing inconsistent retention and deletion semantics.

### 2.4 I/O Models Assume Explicit, Deterministic Invocation

System calls and file I/O presume a program explicitly requests operations. Agents choose tools dynamically. This introduces requirements for tool discovery, tool mediation, and tool attribution.

In a traditional OS, a syscall is allowed or denied based on privileges. For agents, permission must depend on dynamic context, impact radius, and policy constraints. The invocation pipeline must remain deterministic in enforcement and must record not only that an operation occurred but also the lineage that led to it.

### 2.5 Identity and Permissions Assume Stable Principals and Stable Intent

OS security models assume a user identity or service identity with known privileges. Agents act autonomously and may change behavior over time. They often require scoped escalation, revocation, and delegated authority on behalf of humans or services.

A static permission model is insufficient for safe tool use. The system needs capability-based access with explicit scopes, parameter bounds, time limits, and action logging. Capabilities must be revocable and must support least privilege execution environments.

### 2.6 Observability Focuses on Resources, Not Decision Lineage

Traditional observability tracks CPU, memory, disk, network, and process level tracing. Agent systems require decision lineage: what the agent observed, which action it proposed, what policy decided, which tool executed, and what outcome occurred.

Without lineage, operators cannot answer the core operational questions: what did the agent do, why did it do it, and was it allowed. This gap becomes critical in incident response, compliance audits, and safety reviews.

## 3. Definition of an Agent Operating System

This section defines AOS precisely and establishes boundaries to avoid ambiguity about scope, execution, and enforcement.

### 3.1 Core Definition

An Agent Operating System (AOS) is a systems software layer that manages the lifecycle, execution, coordination, and governance of goal-directed agents by extending or reinterpreting classical operating system responsibilities, while preserving deterministic behavior at the system boundary.

This definition separates two properties. Agent internal reasoning may be probabilistic. AOS enforcement and external effects must be deterministic and auditable.

### 3.2 First-class Entities

In a classical OS, first-class entities include processes, threads, file descriptors, and address spaces. In an AOS, additional first-class entities include agent identity, goal and task graphs, capability sets, context state, and execution records.

1. Agent identity is a principal with scoped authority and a defined lifecycle.
2. Goal and task graphs represent intended outcomes and dependencies.
3. Capability sets define which tools and actions are allowed, including scopes and bounds.
4. Context state is the structured memory view presented to the reasoning plane.
5. Execution records are append only audit trails for decisions and actions.

AOS does not replace processes or threads. It interprets them as execution containers beneath the agent abstraction.

### 3.3 Separation of Planes

AOS enforces a separation between reasoning, execution, and policy.

The reasoning plane performs probabilistic inference, planning, and strategy selection. It may be executed locally or remotely and must be treated as untrusted with respect to policy.

The execution plane performs deterministic tool invocation, system calls, and side effects. It enforces policy decisions deterministically and runs actions in least privilege environments.

The policy plane performs authorization, risk checks, compliance constraints, and budget controls. It denies by default where ambiguous and produces auditable decisions with reason codes.

This separation is a core safety property. It allows the system to treat reasoning outputs as proposals rather than commands.

Ambiguity is determined by policy-defined confidence thresholds, incomplete classification state, conflicting governance rules, or inability to establish deterministic authorization semantics for a proposed action.

### 3.4 Trusted Computing Base (TCB)

The AOS architecture implicitly defines a Trusted Computing Base (TCB) consisting of components responsible for deterministic enforcement, mediation, and audit integrity.

Components inside the TCB include the policy engine, capability validation pipeline, mediation layer, execution sandbox controls, audit subsystem, and deterministic state management mechanisms.

Components outside the TCB include probabilistic reasoning systems such as LLM inference engines, retrieval models, summarizers, and heuristic classifiers whose outputs may be incorrect or adversarially influenced.

The architectural objective is not to eliminate nondeterminism from reasoning systems, but to ensure that nondeterministic outputs cannot directly produce side effects without traversing deterministic enforcement paths inside the TCB.

This separation provides a concrete architectural basis for evaluating attack surface, audit integrity, replay guarantees, and enforcement correctness.

Certain higher-level policy evaluations, such as semantic data classification or behavioral anomaly detection, may depend on probabilistic classifiers outside the deterministic enforcement core. In such cases, deterministic enforcement applies to policy outcomes and mediation behavior rather than to the classifier itself.

### 3.5 Determinism at The Boundary

The boundary is the interface between agent outputs and real side effects. Side effects include file writes, process creation, network calls, configuration changes, and external API calls that mutate state.

The boundary is deterministic when authorization decisions are deterministic given inputs, tool invocation parameters are validated deterministically, audit logs are complete and tamper evident, and the mapping from approved action to executed side effect is deterministic.

AOS can tolerate nondeterminism inside reasoning. It cannot tolerate nondeterminism in enforcement.

## 4. AOS Architecture and Components

This section specifies a reference architecture for an Agent Operating System. The intent is to make responsibilities explicit and testable. The reference architecture does not require kernel modifications. It can be realized as user-space services with deterministic enforcement, while leveraging underlying OS primitives for isolation and resource control.

### 4.1 High Level Architecture

The AOS control plane introduces a semantic elevation above process abstractions. Where the classical OS schedules threads, AOS schedules goal progress. Where the classical OS mediates syscalls, AOS mediates tool invocations. This symmetry clarifies that AOS extends OS principles rather than replacing them.

Fig. 1 presents a baseline model of traditional OS mediated execution. Fig. 2 presents the AOS model in which an agentic control plane mediates between probabilistic reasoning and deterministic execution on an underlying OS.

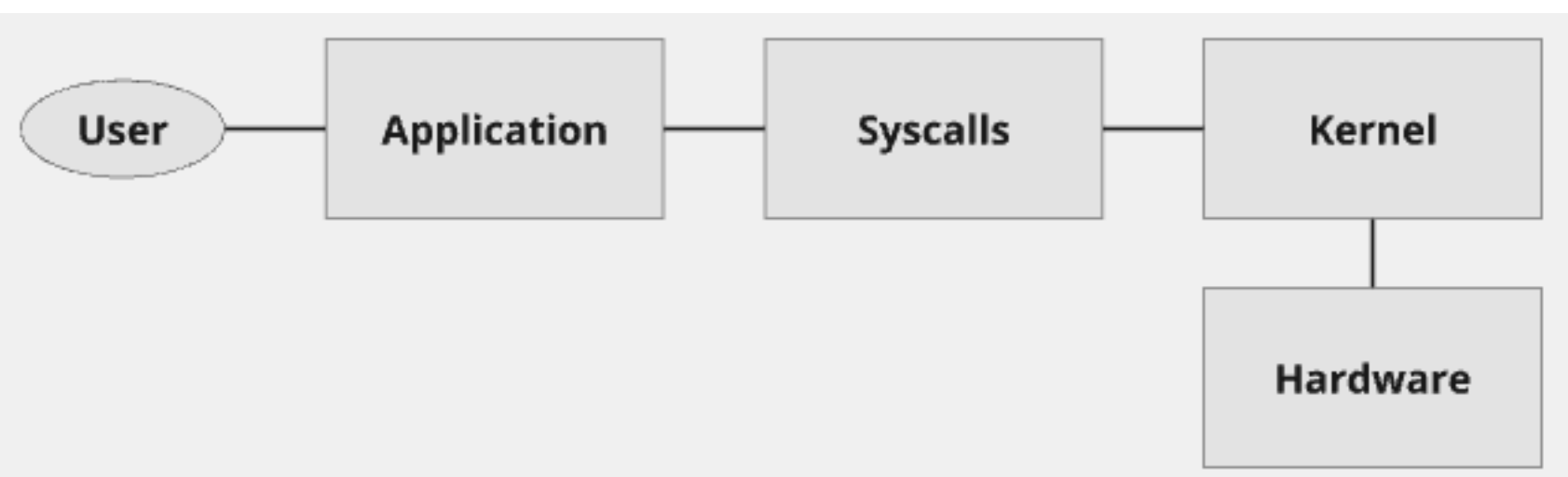

Fig. 1. Classical OS execution model

*A traditional OS mediates hardware via syscalls invoked by deterministic programs. Observability and security decisions are primarily bound to processes and users.*

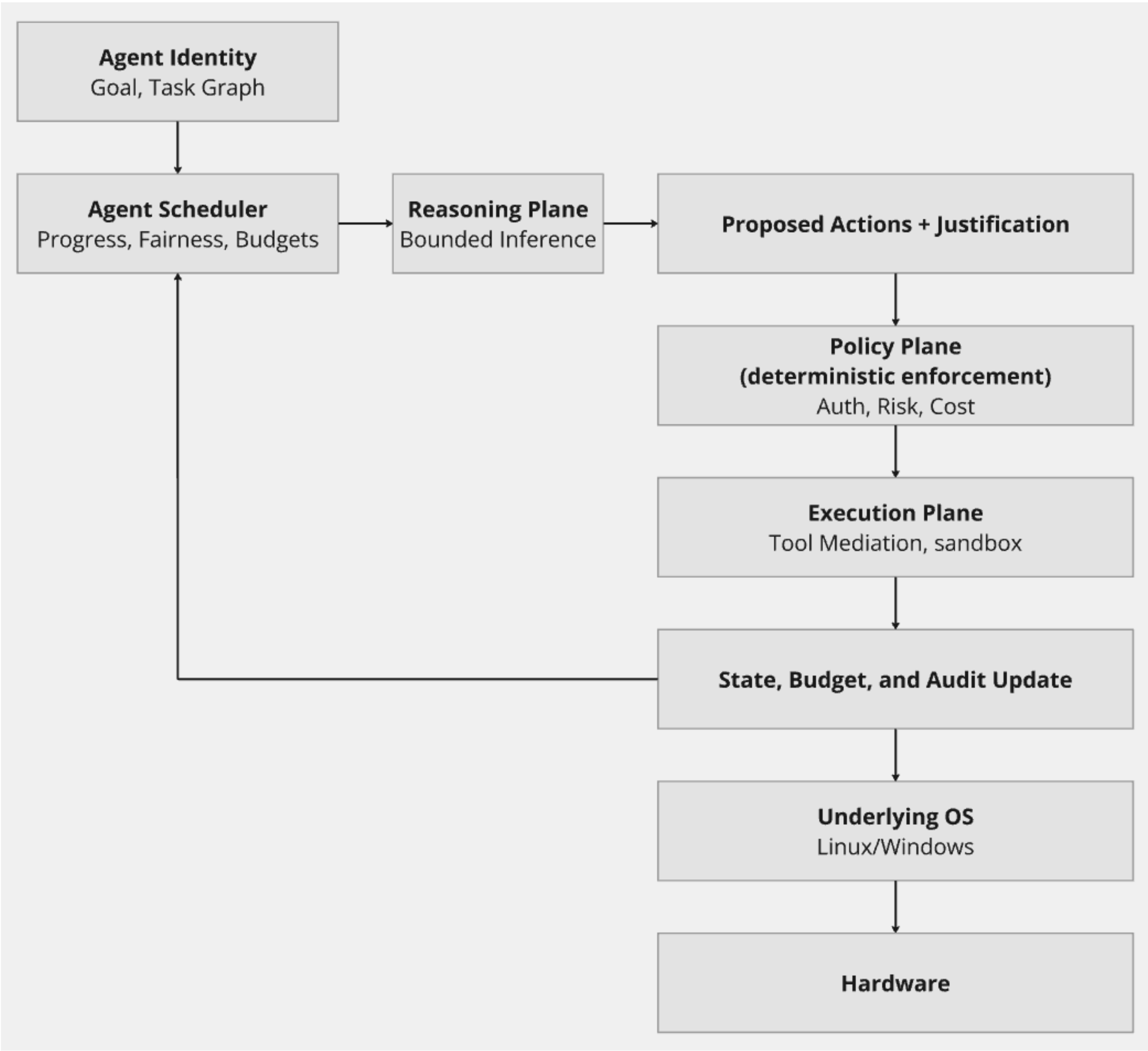

Fig. 2. AOS as an agentic control plane over an OS

*AOS introduces an explicit control plane around agent identity, scheduling, policy enforcement, and mediated execution while relying on the underlying OS for low level resource and isolation primitives.*

### 4.2 Agent Lifecycle Manager

Agents require lifecycle states beyond process states. A representative lifecycle includes created, initialized, active reasoning, awaiting events, executing, suspended, and terminated. The lifecycle manager maintains durable agent identifiers, capability grants and revocations, context store pointers with retention and classification metadata, and deterministic termination semantics.

Termination is nontrivial for agents because they may have outstanding tool calls or delegated tokens. AOS should define deterministic cleanup rules, such as canceling in flight tool requests, revoking tokens, closing sessions, and persisting a final audit closure record. If tool calls are non cancelable, the AOS should record compensating actions or a bounded remediation plan, rather than silently leaving external effects unmanaged.

## 4.3 Agent Scheduler

The agent scheduler is conceptually analogous to CPU schedulers, but its objective function is different. Traditional scheduling optimizes CPU fairness, latency, and throughput. AOS scheduling must optimize goal progress subject to policy, risk, and budget constraints.

Scheduling goals commonly include goal progress per unit cost, responsiveness to urgent events, and fairness across agents and tenants. Scheduling inputs include agent priority and SLA class, cost budgets and rate limits, dependency graph state, risk posture, and behavioral signals such as repeated denials or integrity failures. Scheduling outputs include when to allocate bounded reasoning slices, when to initiate tool calls, when to require human approval, and when to throttle, suspend, or quarantine an agent.

AOS must separate reasoning resources and execution resources. Reasoning slices consume inference time, context construction, and retrieval. Execution slices consume mediated tool time and underlying OS resources. Treating both as a single process risks unbounded inference consumption while appearing idle or low impact at the OS level.

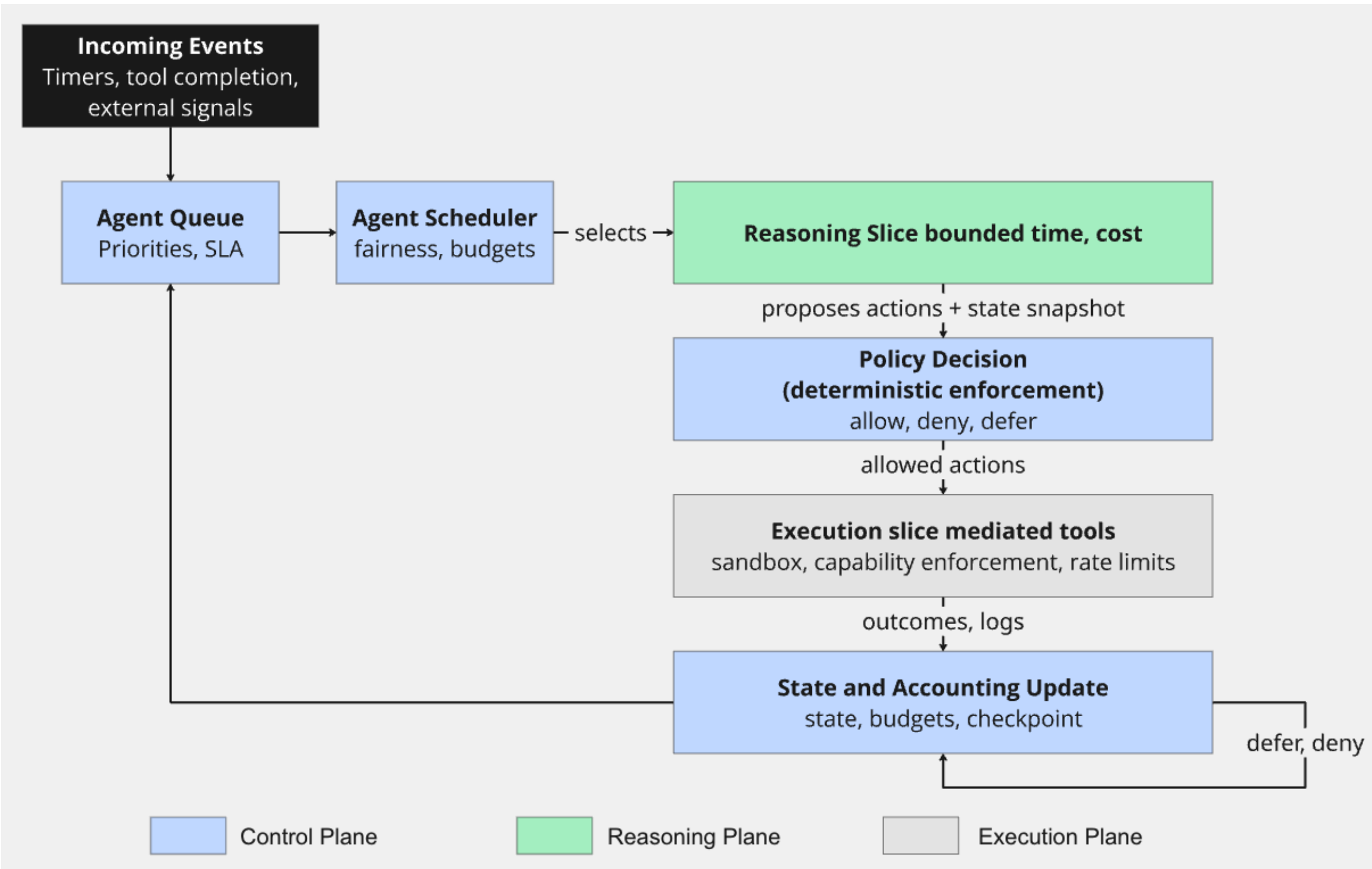

Fig. 3. Scheduling loop with progress checkpoints

*AOS schedules bounded reasoning and execution slices and records progress checkpoints to support operator control, fairness, and auditability.*

## 4.4 Context and Memory Management

AOS treats context and memory as managed resources with explicit semantics, rather than as incidental application state. This is required for auditability, safety, and multi tenant governance.

A useful decomposition is four memory classes. Ephemeral context is a bounded working set constructed for a reasoning slice. Durable agent memory is persistent state across long-lived tasks and must be versioned. Retrieved knowledge is content fetched from documents, tools, or external stores and must carry provenance, integrity metadata, and classification labels. Execution records are append only event streams of actions, policy decisions, and outcomes.

Context construction should be deterministic. Retrieval sets must be recorded. Any summarization that reduces retrieved content into a smaller context representation must record the source inputs and the summarizer version. Filters must be deterministic and logged so that operators can reproduce the effective context view. These properties allow incident response and compliance to reconstruct what the agent could have known at a given moment, without requiring deterministic replay of model internals.

Memory isolation must be explicit. AOS should provide namespaces for memory stores, policy enforced sharing for controlled collaboration, and explicit consent and attribution for shared memory entries. Retention and deletion must be policy driven and recorded. Deletion workflows should preserve evidence that deletion occurred while preventing the content from being reconstructed beyond policy allowances.

### 4.5 Tool and Capability Registry

Tools are mechanisms by which agents cause side effects. In AOS, tools are analogous to system calls, but with explicit mediation. Unlike classical syscalls, however, agent tools may be dynamically discovered, updated, or revoked at runtime. Practical AOS deployments therefore require a privileged admission and classification mechanism that binds dynamically discovered tools to capability scopes, mediation requirements, and audit policies before use. The tool and capability registry is the central contract between agent reasoning and deterministic execution.

A tool definition should include name and version, input schema and validation rules, side effect classification, required capabilities and scopes, rate limits and quotas, deterministic prechecks, post conditions and expected outputs, and audit requirements. This transforms tool invocation from an ad hoc API call into a governed system operation.

Capabilities should be explicit and least privilege. A capability specifies which tools can be invoked, which resource scopes apply, which parameter ranges are permitted, which time windows are valid, and whether human approval is required. Capability checks must be deterministic and deny by default where ambiguous.

Tool invocation should be mediated through a standard pipeline: schema validation and normalization, policy checks including context dependent constraints, risk checks such as sensitive data egress limits, budget checks including cost and rate limits, execution in a least privilege sandbox, post checks on outputs, and append only audit logging with reason codes.

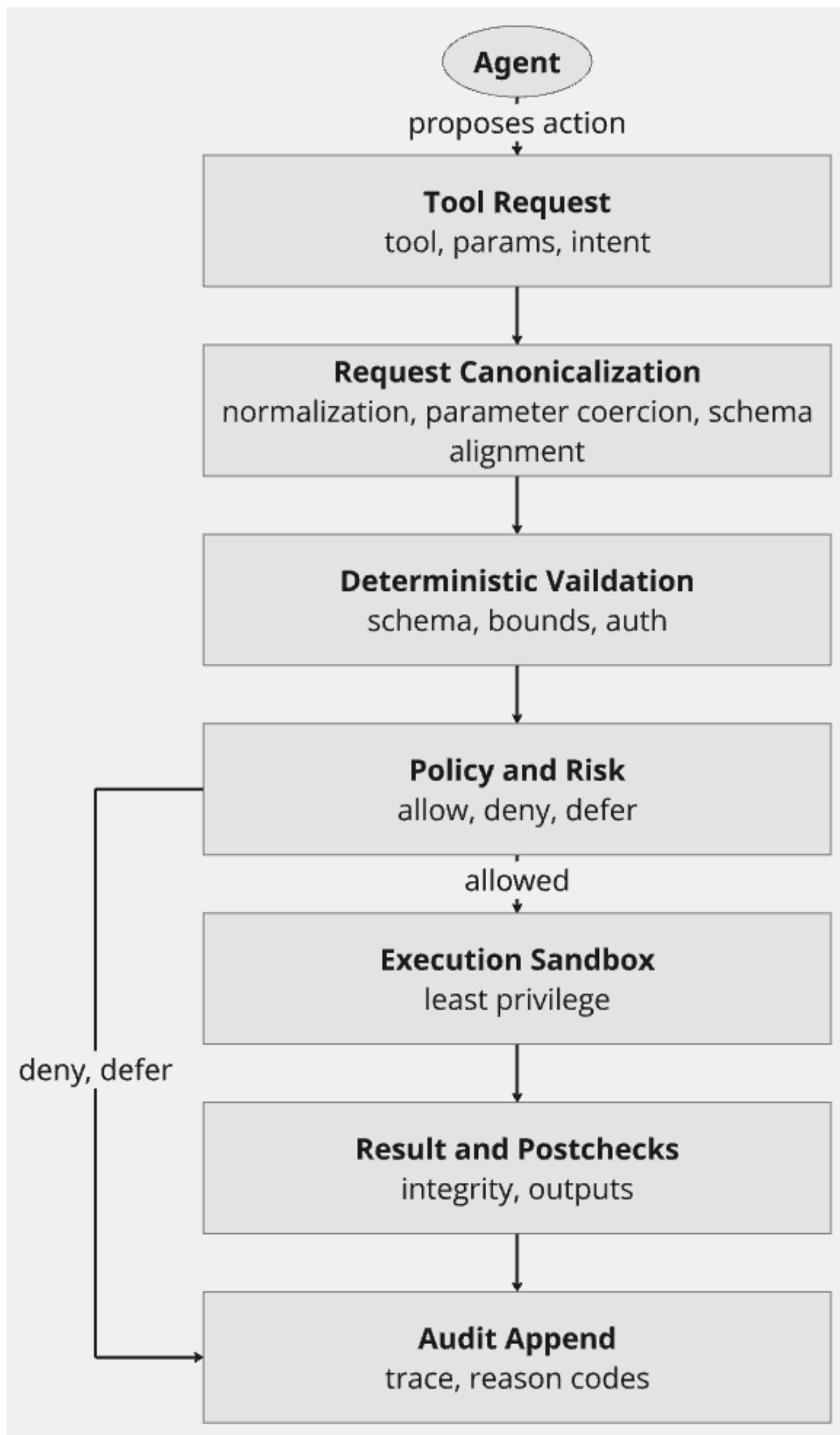

Fig. 4. Tool mediation and deterministic enforcement

*AOS treats tool calls as mediated operations with deterministic checks and complete audit logging. Reasoning proposes. Execution enforces.*

### 4.6 Policy, Permissions, And Trust

AOS policy must handle dynamic context and evolving agent behavior while remaining deterministic. A useful structure is layered policy.

Static authorization binds agent identities to baseline privileges. Capability constraints scope these privileges to specific tools and resources. Context policy constrains actions based on data classification, tenant boundaries, time windows, and environmental signals. Risk policy constrains actions based on side effect classifications, impact radius, novelty, and anomaly indicators. Governance policy introduces approvals, dual control, and escalation for high impact operations.

AOS should support deterministic trust and risk states such as normal, restricted, quarantined, and reviewed. Transitions should be triggered by deterministic signals such as repeated policy denials, suspicious tool patterns, integrity check failures, or unexpected data classification mismatches. The purpose is not to assign opaque trust scores, but to provide a crisp operational posture that operators can understand and audit.

Delegation must be explicit. A user or service may delegate a scope to an agent with expiration, purpose, and audit requirements. Delegation must be revocable and revocation must propagate to tool mediation paths. Actions should be attributed to both the agent identity and the delegating principal.

### 4.7 Observability and Audit

AOS observability must combine system metrics with decision lineage. Traditional OS telemetry answers what resources were used. AOS telemetry must also answer why actions were taken and whether they were allowed.

A minimum observability stack includes resource metrics, agent progress metrics such as goal state and blocked reasons, decision traces with proposed actions and policy decisions, tool traces including inputs and outputs, and data lineage that records which data influenced actions and how it was classified.

Audit records should be structured and append only. At minimum, each action record should include agent and tenant identifiers, timestamps and monotonic sequence identifiers, goal or task identifiers, input summaries with references to detailed inputs, proposed action parameters, policy decision reason codes, tool invocation metadata including environment constraints, output summaries with integrity checks, and links to underlying OS traces.

Deterministic replay should focus on enforcement pipelines. Full deterministic replay of probabilistic reasoning may be impractical. However, AOS should enable deterministic re-evaluation of policy decisions and deterministic reconstruction of context views from recorded retrieval sets and filters. This requires complete capture of all inputs that materially influenced the enforcement decision at the time of evaluation, including retrieved artifacts, summarizer outputs, classifier outputs, policy versions, environmental context, and mediation parameters. Replay guarantees are therefore bounded by completeness of capture rather than deterministic reproduction of probabilistic reasoning internals.

### 4.8 Architectural Invariants

- Invariant 1 - No side-effecting action is executed without a deterministic policy decision of allow.
- Invariant 2 - All policy outcomes (allow, deny, defer) are recorded in an append-only audit record prior to rescheduling.
- Invariant 3 - Scheduling decisions depend only on observable state and budgets, not on internal reasoning tokens.
- Invariant 4 - Underlying OS remains the sole mediator of hardware resources.

## 5. Integration Models

AOS can be implemented in multiple ways. The appropriate model depends on risk tolerance, performance constraints, deployment environments, and the required strength of enforcement. This section evaluates several integration models and highlights tradeoffs.

### 5.1 AOS As a User-space Runtime

In this model, AOS runs as user-space services and libraries while the underlying OS remains unchanged. The agent runtime process manages agent lifecycle and reasoning workers. A policy engine runs as a service that produces deterministic authorization and risk decisions. Tool mediation is implemented as a sidecar or gateway that all tool traffic must traverse. Observability integrates with existing logging, tracing, and audit sinks, while adding decision lineage semantics.

Advantages include low risk adoption, rapid iteration without kernel changes, and portability across Linux and Windows. Existing isolation boundaries such as processes, containers, and virtual machines can be reused. Limitations include limited visibility into kernel scheduling decisions and the risk of bypass if agents can access tools or network egress outside mediation. Strong deployments therefore combine AOS mediation with OS and network controls that force all side effects through controlled channels.

### 5.2 AOS As an OS Extension

In this model, AOS introduces OS level primitives or hooks. Potential primitives include agent identity as an OS recognized principal, capability tokens integrated with syscall or object access layers, scheduler hooks that treat agent progress as a first-class signal, and kernel level audit streams with richer attribution.

The primary advantage is stronger enforcement because the OS mediates at the syscall and object access boundary. This reduces bypass opportunities and can improve observability. The primary risks are increased kernel complexity, expanded attack surface, reduced portability, and slower evolution. A pragmatic intermediate approach is to leverage existing hook mechanisms such as Linux security modules, eBPF, and Windows policy frameworks to implement selective enforcement rather than introducing new kernel subsystems.

### 5.3 AOS As a Distributed Control Plane

In this model, AOS is a cluster control plane that maintains agent identity, goals, policy, and audit centrally, while delegating execution to node level executors in restricted sandboxes. This model is motivated by distributed agent workloads that coordinate across services and data sources, and by organization level governance requirements that cannot be expressed per machine.

AOS overlaps with orchestrators on scheduling and placement but differs in several key ways. It schedules based on goal progress and policy constraints rather than on static resource requests alone. It defines deterministic policy boundaries for agent actions and standardizes decision lineage as a first-class output. It treats tool mediation as a governed execution path rather than as an application library concern. In practice, an AOS control plane can be implemented on top of existing cluster managers, but it introduces additional semantic layers for agent identity, capabilities, and audit.

### 5.4 AOS Subsuming Parts of the OS

A long horizon model is that AOS subsumes selected OS responsibilities over time, particularly at higher layers of execution and governance. Candidate responsibilities include higher level scheduling of tasks and tool invocations, policy driven access control beyond static permission models, and structured audit and lineage interfaces that become primary operational surfaces.

Core kernel responsibilities should remain OS native, including memory paging, low level CPU scheduling primitives, hardware abstraction, driver stacks, and file systems. From a systems perspective, the credible path is for AOS to evolve as a deterministic control plane that increasingly governs side effects and policy, while the kernel continues to provide foundational resource and isolation primitives.

## 6. Mapping To Existing OSes

This section grounds AOS in real primitives and shows how AOS responsibilities can be realized on Linux and Windows without requiring full OS replacement. The goal is to provide an actionable migration path for practitioners and a concrete discussion surface for OS researchers.

### 6.1 Mapping to Linux

Linux already provides mechanisms that can host AOS components. Agents can run as processes or containers. Isolation can be expressed via namespaces for process identifiers, mounts, networking, and users. Resource bounds can be enforced via cgroups for CPU, memory, and I/O.

AOS agent scheduling can be approximated by placing each agent or agent class into dedicated cgroups and controlling CPU shares, quotas, and priorities. Event driven activation and bounded inference bursts reduce thrashing. Reasoning workers and tool executors should run in separate sandboxes to preserve least privilege. For example, an inference worker may require access to memory stores but no network egress, while a tool executor may require network access to a specific destination but minimal filesystem access.

Tool mediation and deterministic enforcement can leverage Linux primitives: seccomp filters to restrict syscalls, mandatory access control via AppArmor or SELinux, controlled egress via network namespaces and firewall rules, and policy enforcement via eBPF programs for network and syscall level observability. AOS must propagate stable agent action identifiers through execution so that kernel level traces can be correlated with decision lineage.

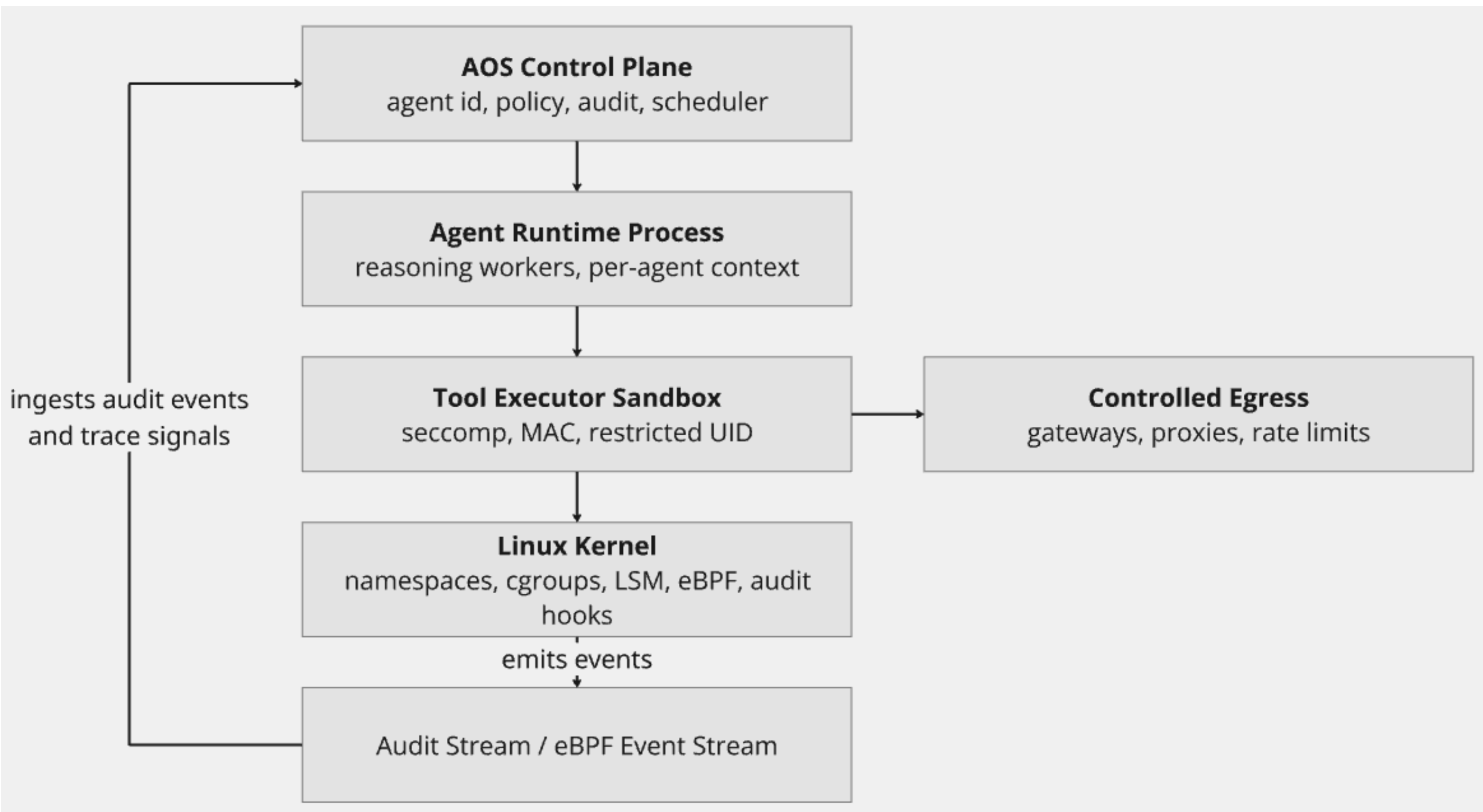

Fig. 5. Linux mapping overview

*AOS components can be realized on Linux using cgroups, namespaces, seccomp, mandatory access control, and eBPF for deterministic enforcement and observability. Controlled egress reduces bypass opportunities.*

## 6.2 Mapping to Windows

Windows provides different primitives, with both opportunities and constraints. Agents can run as services or scheduled tasks and can be isolated via Windows containers or Hyper V isolation where appropriate. Resource controls can be applied using job objects and container resource limits.

Windows security is centered on access tokens and security descriptors. This maps well to scoped delegation if used carefully. AOS can issue restricted tokens for tool execution, strip privileges for least privilege, and enforce object level access through access control lists. Tool execution should be brokered through controlled processes that hold the minimal required privileges and that emit deterministic audit logs.

Observability can leverage Event Tracing for Windows (ETW) and centralized logging. AOS should correlate ETW events with agent action identifiers and persist immutable audit records that include policy decisions and reason codes. Like Linux, the key requirement is end to end correlation between agent level lineage and OS level events.

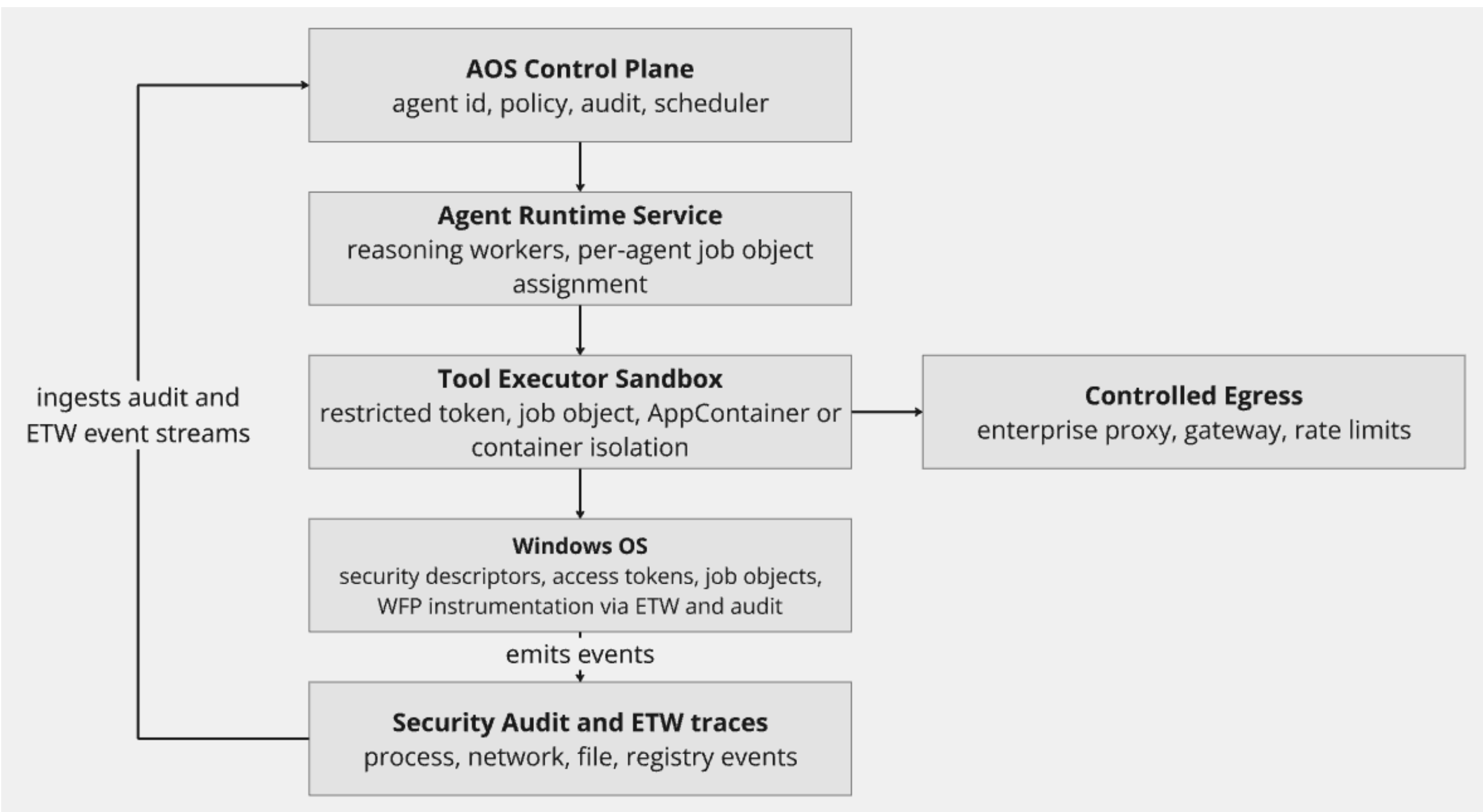

Fig. 6. Windows mapping overview

*On Windows, AOS can leverage service models, restricted security tokens, job objects, and ETW to implement deterministic enforcement and traceability, typically through brokered tool execution.*

## 6.3 Comparative Analysis: Linux and Windows For AOS

Linux strengths include flexible low level isolation primitives, composable enforcement via mandatory access control and seccomp, and powerful observability hooks via eBPF. Windows strengths include strong identity and access control mechanisms via tokens and descriptors, mature centralized logging and enterprise policy controls, and consistent service management.

A shared challenge is that neither OS natively models agent intent, goal progress, or decision lineage. AOS must provide that semantic layer and propagate stable identifiers through execution paths so that kernel and OS events can be attributed to agent actions.

## 7. Linux Vs Windows Subsystem Comparison For AOS

This section compares Linux and Windows as substrates for implementing AOS responsibilities. The objective is not to rank operating systems broadly, but to identify concrete primitives that enable deterministic enforcement, isolation, observability, and delegated execution for agent workloads. For each subsystem we list (a) relevant primitives, (b) what they enable for AOS, and (c) limitations or gaps that AOS must compensate for.

### 7.1 Identity, Principals, And Delegation

Linux primitives

1. UID, GID, supplementary groups
2. Capabilities (fine grained privileges such as CAP_NET_ADMIN)
3. Mandatory access control frameworks (SELinux, AppArmor)
4. PAM and pluggable authentication
5. Namespaces, particularly user namespaces for remapping identities

AOS relevance

1. Map agent identity to a stable principal for ownership and attribution, while keeping agent level identity in AOS metadata.
2. Use capabilities and MAC to implement least privilege for tool execution sandboxes.
3. Use user namespaces to isolate tool execution and prevent accidental privilege inheritance.

Linux limitations for AOS

1. OS identity is still primarily user and process oriented. It does not model “delegation for a specific goal” as a first-class concept.
2. Fine grained constraints over parameters of an action are not naturally expressed at the OS identity layer. This pushes decision logic into AOS policy and mediation.
3. Attribution across chains of processes requires careful propagation of correlation identifiers.

Windows primitives

1. Security identifiers (SIDs) and access tokens
2. Privileges within tokens (e.g., SeDebugPrivilege)
3. Access control lists (DACL, SACL) and security descriptors
4. User Account Control and integrity levels
5. Service accounts and managed service identities in enterprise deployments

AOS relevance

1. Restricted tokens support strong least privilege execution for tools and side effects.
2. ACLs allow object level enforcement, useful for file and registry access boundaries for agents.
3. Auditing via SACL and centralized event pipelines support governance.

Windows limitations for AOS

1. As with Linux, “goal scoped delegation” is not a native OS primitive. It must be implemented by AOS as policy and token issuance discipline.
2. Ensuring every agent side effect flows through a mediated broker is an architectural requirement. The OS will not guarantee this unless tool execution is strictly constrained.

Key AOS design implication: AOS should define delegation as an AOS level construct that results in OS level principals and tokens, but remains auditable and revocable at the AOS plane.

## 7.2 Isolation and Sandboxing

Linux primitives

1. Namespaces: pid, mount, net, ipc, uts, user
2. cgroups v1 and v2 (CPU, memory, I/O, pids)
3. seccomp for syscall filtering
4. Linux capabilities and no new privileges flag
5. MAC enforcement: SELinux, AppArmor
6. Containers built from these primitives

AOS relevance

1. Run tool execution in sandboxes with minimal syscalls, limited filesystem views, restricted network, and bounded resources.
2. Enforce deterministic containment rules independent of model behavior.
3. Limit blast radius of injected or erroneous tool requests.

Linux limitations for AOS

1. seccomp and MAC policy creation is complex and error prone. AOS must provide policy generation, testing, and safe defaults.
2. Container escape risks exist if policies are misconfigured, so mediation must be defense in depth.
3. Enforcing “only mediated tools” requires additional constraints at network and API layers.

Windows primitives

1. Job objects (resource constraints and process grouping)
2. Windows containers and Hyper V isolation
3. AppContainer and capability-based sandboxing for certain app models
4. Code integrity and application control policies
5. Windows Defender and enterprise controls for execution policy

AOS relevance

1. Group tool processes in job objects to enforce CPU and memory caps and enable termination semantics.
2. Use Windows containers or Hyper V isolation for high-risk tools.
3. Rely on restricted tokens and application control to limit what can execute.

Windows limitations for AOS

1. Sandboxing models vary by subsystem and app type. AOS must standardize how tools are packaged and executed.
2. Some constraints are easier at the enterprise policy layer than at the OS kernel layer, which increases dependence on environment configuration.

Key AOS design implication: Tool execution should be treated as a controlled workload class with standardized packaging and sandbox profiles per tool category, irrespective of OS.

## 7.3 Scheduling and Resource Management

Linux primitives

1. CFS scheduler, real time scheduling classes
2. nice values and priorities
3. cgroups CPU quotas and shares
4. I/O controllers in cgroups v2
5. OOM killer and memory pressure signals

AOS relevance

1. Enforce bounded reasoning and execution budgets using cgroups for inference workers and tool sandboxes.
2. Use separate resource pools for reasoning plane and execution plane to avoid hidden starvation.
3. Use memory pressure signals to trigger deterministic degradation such as shrinking context windows or pausing low priority agents.

Linux limitations for AOS

1. The kernel schedules threads, not goals. Goal progress must be tracked and acted upon by AOS.
2. Distinguishing “useful reasoning” from “thrashing” is not visible to the OS. AOS must impose time slice limits and checkpointing.
3. QoS across agents is difficult if all inference runs in a shared runtime process. AOS should isolate per agent reasoning workers where feasible.

Windows primitives

1. Windows scheduler priorities and quantum management
2. Job objects for CPU rate control
3. Resource controls in containers
4. I/O prioritization and system resource manager features in enterprise environments

AOS relevance

1. Like Linux, use job objects and container controls to bound inference and tool execution.
2. Use OS performance counters and ETW signals to detect pressure and enforce throttles.

Windows limitations for AOS

1. Fine grained control is often mediated by enterprise policy and management layers.
2. As with Linux, the OS does not understand goal progress. AOS must implement checkpoints and progress accounting.

Key AOS design implication: AOS must introduce an agent scheduler that produces resource intents, while the OS enforces low level allocation via cgroups or job objects. The AOS scheduler remains the source of truth for fairness and progress.

## 7.4 I/O, Networking, And Controlled Egress

Linux primitives

1. netfilter, nftables, iptables for policy enforcement

2. Network namespaces for isolation
3. eBPF based traffic accounting and policy hooks
4. Proxy and gateway patterns at user-space
5. Filesystem controls via mount namespaces and read only overlays

AOS relevance

1. Ensure that all external calls from tool sandboxes are routed through controlled egress gateways for logging and policy enforcement.
2. Enforce deterministic allowlists, rate limits, and data loss prevention rules at the boundary.

Linux limitations for AOS

1. Application-level policies such as per API call auditing require cooperation from mediation layers, not only kernel filters.
2. Strictly preventing direct outbound calls without a proxy is feasible but operationally delicate and must be tested.

Windows primitives

1. Windows Firewall and enterprise network policy
2. Proxy settings, enterprise gateways
3. Windows Filtering Platform for network filtering and inspection
4. ETW network events

AOS relevance

1. Similar controlled egress design, with OS policy enforcement plus brokered tool calls.
2. Centralized enterprise policy can support strong governance if configured.

Windows limitations for AOS

1. Consistency depends heavily on enterprise configuration and managed endpoints.
2. Some applications may bypass proxy settings, requiring stronger broker and sandbox constraints.

Key AOS design implication: Treat network as a side-effecting tool category. Require mediated egress by default and treat direct outbound networking from agent sandboxes as a policy exception.

## 7.5 Observability, Auditing, And Trace Correlation

Linux primitives

1. eBPF for syscall, network, and scheduler tracing
2. auditd subsystem
3. perf and ftrace for performance
4. journald, syslog
5. cgroup metrics and procfs

AOS relevance

1. Correlate OS level events with AOS action ids to form decision lineage.
2. Use eBPF and audit logs to detect bypass attempts and unexpected syscalls.

3. Record resource usage per agent and per tool for evaluation and billing like accounting.

Linux limitations for AOS

1. Correlation is not automatic. AOS must propagate stable correlation identifiers into tool processes and logs.
2. Capturing full decision lineage requires AOS controlled logging of reasoning inputs and outputs, with careful privacy controls.

Windows primitives

1. Event Tracing for Windows (ETW)
2. Windows Event Log and auditing policies
3. Performance counters
4. Sysmon in enterprise contexts
5. Service control manager events

AOS relevance

1. ETW supports high fidelity event pipelines that can be stitched into AOS audit records.
2. Strong enterprise audit posture is feasible when policy is correctly configured.

Windows limitations for AOS

1. As with Linux, AOS must ensure correlation id propagation and consistent tool brokering.
2. High volume event streams require careful sampling and retention policies, and AOS should define which events are mandatory for compliance.

Key AOS design implication: AOS must define a canonical audit schema and correlation protocol. OS observability enriches but does not replace AOS decision logs.

## 8. Security, Safety, And Governance Implications

Agents introduce new threat surfaces and new failure modes. This section defines explicit threat models and deterministic mitigations that align with established OS security principles.

### 8.1 Threat Model Categories

AOS must consider at least the following threat categories.

1. Prompt and context manipulation: An adversary attempts to influence agent reasoning by injecting malicious instructions into inputs or retrieved memory.
2. Tool misuse: An agent invokes tools with harmful parameters or unintended side effects.
3. Privilege escalation: The agent obtains broader access than intended through misconfigured delegation or tool chaining.
4. Data exfiltration: The agent leaks sensitive data through tool outputs or uncontrolled network calls.
5. Supply chain and tool integrity: Tools are compromised or return malicious or misleading outputs.
6. Audit evasion: The agent bypasses mediation to avoid logging or enforcement.
7. Control plane compromise: Adversaries compromise the AOS control plane itself, including policy engines, capability stores, mediation services, or audit infrastructure, thereby bypassing enforcement guarantees.

## 8.2 Stateful and Compositional Attacks

Many real-world agent failures are compositional rather than atomic. Individually permitted actions may collectively produce harmful outcomes over time, including gradual configuration weakening, incremental data exfiltration, or multi-step privilege expansion. Per-action deterministic enforcement is necessary but not sufficient to mitigate such attacks. Practical AOS deployments may therefore require stateful policy evaluation over action histories, cumulative behavioral patterns, and longitudinal context. This introduces additional challenges because stateful anomaly detection and behavioral classification may themselves rely on probabilistic mechanisms outside the deterministic enforcement core.

## 8.3 Deterministic Enforcement Principles

Deterministic enforcement principles are needed to ensure controllability regardless of reasoning quality.

1. Deny by default for side-effecting tools unless explicitly allowed by capability.
2. Capabilities must be scoped, time bounded, and revocable.
3. All tool calls must flow through mediation. This must be enforced by OS and network controls, not by convention.
4. Policy decisions must produce deterministic reason codes and be logged.
5. Output filters for sensitive data classes must be deterministic and applied after tool execution but before external egress where applicable.
6. Safe failure modes must prefer denial of side effects under policy engine outages or audit sink failures.

## 8.4 Prompt Injection as An Untrusted Input Problem

Prompt injection should be treated as a form of untrusted input, analogous to user supplied data in traditional OS security models. The mitigation is not to assume that reasoning can be made perfectly robust. The mitigation is to constrain execution through deterministic mediation.

Model outputs should be treated as proposals rather than commands. Each proposed action must pass deterministic policy checks and capability validation before execution. Tools should run in least privilege sandboxes. The AOS must prevent the agent from turning injected text into privileged side effects without passing through enforcement points.

## 8.5 Human in The Loop Governance

Some operations require approvals due to impact radius or regulatory requirements. AOS should support approval gates based on tool classification and change impact, and it should produce approval requests that are understandable and auditable. Approvals should be time bounded and scope limited. They are not substitutes for deterministic policy. They are governance mechanisms for high impact operations within a deterministic enforcement framework.

## 8.6 Bounded Autonomy Levels

AOS deployments benefit from explicit autonomy levels that align with enterprise change management.

1. Read only. No side effects. Tools limited to observation and retrieval.
2. Low risk. Limited and reversible side effects under strict scopes.
3. Controlled. Broader actions with full policy, audit, and anomaly monitoring.
4. High impact. Actions require approvals or dual control and may require additional verification steps.

These levels allow operators to match agent autonomy to system criticality and compliance constraints.

## 9. Related Work and Comparison

This section situates Agent Operating Systems relative to established systems mechanisms. The comparison emphasizes what those mechanisms solve, what they intentionally do not solve, and where AOS introduces new requirements, particularly around deterministic enforcement of tool use, decision lineage, and goal-oriented scheduling.

### 9.1 Containers and OS Level Virtualization

What containers provide

1. Process level isolation using namespaces and resource controls
2. Reproducible packaging of dependencies and runtime environments
3. Operational primitives for deployment, scaling, and lifecycle management

What containers do not provide for agent systems

1. Agent identity as a principal with delegated authority and revocation semantics
2. Deterministic mediation for side-effecting actions beyond OS syscalls
3. Decision lineage, including why an action was taken and under which policy constraints
4. Goal progress scheduling, since containers run workloads rather than manage intent

Relationship to AOS: AOS can use containers as an execution substrate for tools and agent runtimes. In an AOS design, containers become a containment mechanism, while AOS provides the control plane for goal scheduling, tool mediation, and audit. Containers help with isolation. They do not define the semantics of agent autonomy.

### 9.2 Hypervisors and Virtual Machines

What hypervisors provide

1. Strong isolation at hardware boundaries
2. Independent OS instances per workload
3. Reduced blast radius for compromised kernels or containers

What hypervisors do not provide for agent systems

1. Fine grained capability mediation for tool invocation within a VM
2. Cross VM policy enforcement tied to agent identity and delegated authority
3. Unified decision lineage across reasoning and execution components

Relationship to AOS: Hypervisors are useful for high-risk tool sandboxes or strict tenant separation. AOS can schedule tool execution into VMs for stronger isolation. However, hypervisors address isolation, not governance. AOS must still control what actions execute and produce the audit trail.

### 9.3 Cluster Schedulers and Kubernetes Style Orchestration

What cluster orchestration provides

1. Declarative desired state for workloads
2. Scheduling based on resource requests and node availability
3. Lifecycle automation and self-healing at the service level

4. Namespace isolation and multi tenancy controls

What orchestration does not provide for agent systems

1. Goal progress and intent scheduling. Orchestrators schedule pods and containers, not goals and plans.
2. Deterministic tool mediation. Orchestrators do not enforce per action parameter constraints or reason codes.
3. Decision lineage. Orchestrators explain placement and lifecycle events, not autonomous action selection within workloads.
4. Agent specific governance such as approvals, delegated authority, and bounded autonomy levels.

Relationship to AOS: AOS as a distributed control plane can be implemented on top of Kubernetes. In that model, Kubernetes manages where components run, while AOS governs what actions occur and under what policy. This separation mirrors a common system layering: infrastructure orchestration handles placement, while the control plane handles intent, policy, and audit.

## 9.4 Policy Engines and Authorization Systems

What policy engines provide

1. Declarative authorization logic
2. Centralized policy management and auditing hooks
3. Consistent decisions across services

What policy engines do not provide for agent systems

1. End to end action mediation. Policy engines decide to allow or deny, but they do not guarantee execution through a mediated sandbox.
2. Tight coupling to resource accounting and budgets across reasoning and execution phases.
3. Built in decision lineage that includes the reasoning context and proposed action chain, unless integrated explicitly.

Relationship to AOS: Policy engines naturally belong in the AOS policy plane. However, AOS must extend beyond authorization to include deterministic enforcement, tool brokering, and trace correlation. In other words, policy engines can be one component of AOS, but they are not sufficient to define an AOS.

## 9.5 Agent Frameworks and Tool Calling Runtimes

General observation: Many agent runtimes provide planning loops, tool calling, and memory abstractions at the application layer. These systems often prioritize developer ergonomics and rapid iteration.

Gap addressed by AOS: AOS focuses on systems responsibilities that application frameworks typically treat as optional:

1. Deterministic enforcement boundaries
2. Capability based tool access with revocation
3. Auditability and operator comprehensibility
4. Resource scheduling and isolation across many agents

Relationship to AOS: AOS is complementary. Frameworks can run within the reasoning plane, while AOS constrains execution and provides governance. This separation reduces the requirement that application code be perfectly safe, since enforcement is centralized and deterministic.

## 10. Evaluation Criteria and Open Research Problems

Evaluation of AOS should focus on systems properties rather than on model accuracy. The core question is whether the system remains deterministic, auditable, and secure under probabilistic reasoning and adversarial conditions.

### 10.1 Evaluation Criteria

Key criteria include deterministic enforcement correctness, audit completeness, containment and least privilege, operational comprehensibility, bounded performance overhead, and safe failure modes.

Deterministic enforcement correctness requires that the same inputs produce the same authorization outcomes and that policy decisions are explainable via logged reason codes. Audit completeness requires that every side-effecting action is recorded with lineage and that logs are tamper evident and queryable. Containment requires that tools cannot be invoked outside mediation and that capability scopes are minimal. Operational comprehensibility requires that operators can understand agent state, blocked reasons, and actions without interpreting model internals. Performance overhead should be bounded and measurable. Failure modes should degrade safely, typically by denying side effects when policy or audit infrastructure is unavailable.

### 10.2 Benchmark Scenarios

Representative benchmark scenarios include:

1. Tool heavy workflow. Many tool calls with mixed read and write operations to measure mediation latency and audit coverage.
2. Multi agent coordination. Several agents sharing data under strict policy to measure isolation, controlled sharing, and deadlock or starvation avoidance.
3. Adversarial input campaign. Injection attempts through documents and messages to measure containment and policy resilience.
4. Incident response replay. Reconstruction of actions and policy decisions to measure trace usability and deterministic reconstruction of context views.

### 10.3 Open Research Problems

Open problems include agent scheduling theory for goal progress and fairness, memory semantics for deterministic context construction with provenance, formal verification boundaries for enforcement pipelines, deterministic trust and risk state transitions that avoid opaque scoring, and cross system identity and delegation standards that ensure revocation and attribution across tool chains.

## 11. Conclusion And Future Work

Agentic systems alter the computational unit from deterministic programs to goal-directed entities with probabilistic reasoning and dynamic tool use. Existing operating systems provide strong low-level primitives for isolation and hardware mediation but do not model intent, decision lineage, or deterministic governance at the boundary where autonomous actions create real side effects.

This paper defined an Agent Operating System as a control plane that makes agents first-class entities while preserving deterministic enforcement, auditability, and security. We decomposed AOS into concrete components, evaluated integration models, and mapped AOS concepts to Linux and Windows primitives. Immediate practical paths emphasize user-space AOS runtimes with strong mediation supported by OS enforcement mechanisms. Longer term work includes formal scheduling models for agents, memory semantics for context construction and provenance, and verifiable enforcement pipelines that make deterministic governance tractable at scale.

Multi-agent delegation chains, cascading revocation semantics, and policy arbitration across cooperating agents remain open research problems beyond the scope of this paper.

## 12. References

[1] A. S. Tanenbaum and H. Bos, *Modern Operating Systems*, 4th ed., Pearson, 2014.
[2] A. Silberschatz, P. B. Galvin, and G. Gagne, *Operating System Concepts*, 10th ed., Wiley, 2018.
[3] M. Kerrisk, *The Linux Programming Interface*, No Starch Press, 2010.
[4] P. Barham et al., “Xen and the art of virtualization,” *Proceedings of the 19th ACM Symposium on Operating Systems Principles (SOSP)*, 2003.
[5] B. Burns, B. Grant, D. Oppenheimer, E. Brewer, and J. Wilkes, “Borg, Omega, and Kubernetes,” *Communications of the ACM*, vol. 59, no. 5, pp. 50–57, 2016.
[6] D. Merkel, “Docker: Lightweight Linux containers for consistent development and deployment,” *Linux Journal*, vol. 2014, no. 239, 2014.
[7] J. Edge, “A deep dive into cgroups,” *LWN.net*, 2016.
[8] P. Loscocco and S. Smalley, “Integrating flexible support for security policies into the Linux operating system,” *Proceedings of the USENIX Annual Technical Conference*, 2001.
[9] M. Russinovich, D. Solomon, and A. Ionescu, *Windows Internals*, 7th ed., Microsoft Press, 2017.
[10] H. Levy, *Capability-Based Computer Systems*, Digital Press, 1984.
[11] A. Starovoitov and D. Borkmann, “eBPF – In-kernel virtual machine,” Linux Foundation Documentation, 2015.
[12] T. Brown et al., “Language models are few-shot learners,” *Advances in Neural Information Processing Systems (NeurIPS)*, 2020.